# Observation of magnetic domain and bubble structures in magnetoelectric $Sr_3Co_2Fe_{24}O_{41}$

H. Nakajima[1,2], H. Kawase[3], K. Kurushima[4], A. Kotani[1], T. Kimura[5], and S. Mori[1]

[1]*Department of Materials Science, Osaka Prefecture University, Sakai, Osaka 599-8531, Japan*

[2]*Department of Applied Quantum Physics and Nuclear Engineering, Kyushu University, Fukuoka 819-0395, Japan*

[3]*Division of Materials Physics, Graduate School of Engineering Science, Osaka University, Toyonaka, Osaka, 560-8531, Japan*

[4]*Toray Research Center, Ohtsu, Shiga 520-8567, Japan*

[5]*Department of Advanced Materials Science, University of Tokyo, Kashiwa, Chiba 277-8561, Japan*

The magnetic domain and bubble structures in the Z-type hexaferrite $Sr_3Co_2Fe_{24}O_{41}$ were investigated using Lorentz microscopy. This hexaferrite exhibits a room-temperature magnetoelectric effect that is attributed to its transverse conical spin structure (TC phase). Upon heating, the TC phase transforms into a ferrimagnetic phase with magnetic moments in the hexagonal *ab* plane between 410 and 480 K (FM2 phase) and into another ferrimagnetic phase with moments parallel to the *c* axis between 490 and 680 K (FM1 phase). Accordingly, in this study, the magnetic domain structures in $Sr_3Co_2Fe_{24}O_{41}$ were observed to change dramatically with temperature. In the TC phase, irregular fine magnetic domains were observed after cooling the specimen from the FM2 to TC phase. In the FM1 phase, striped magnetic domain walls with pairs of bright and dark contrast were formed parallel to the *c* axis. Upon applying an external magnetic field, the striped magnetic domain walls transformed into magnetic bubbles. The topology of the magnetic bubbles was dependent on the angle between the external magnetic field (*H*) direction and the easy *c* axis. Namely, magnetic bubbles with the topological number $N = 1$ (type I) were created for *H*//*c*, whereas magnetic bubbles with $N = 0$ (type II) were created when the magnetic field was tilted from the *c* axis by 5°. We attribute the high magnetocrystalline anisotropy of $Sr_3Co_2Fe_{24}O_{41}$ to the emergence of magnetic bubbles in the FM1 phase.

## I. INTRODUCTION

Magnetoelectric (ME) materials have been the subject of intensive studies from both scientific and technological viewpoints because their ferroelectricity (magnetism) can be controlled with the application of a magnetic (electric) field [1]. Helimagnets are of particular interest because they exhibit gigantic ME responses due to the inverse Dzyaloshinskii–Moriya (DM) interaction or spin–current mechanism [2, 3]. Among this class of materials, hexagonal ferrites (hexaferrites) showing helimagnetic orders are promising candidates for future spintronic devices and multibit memory because many hexaferrites exhibit the ME effect, and some of them exhibit a significant ME effect with the application of relatively low magnetic fields [4]. For example, Y-type hexaferrites, such as $Ba_{0.5}Sr_{1.5}Zn_2Fe_{12}O_{22}$ and $Ba_2Ni_2Fe_{12}O_{22}$, and M-type hexaferrites, such as $BaFe_{12-x-0.05}Sc_xMg_{0.05}O_{19}$, exhibit proper screw or longitudinal conical structures in zero magnetic field, and upon application of a magnetic field, they exhibit ferroelectricity and the ME effect via the spin–current mechanism [5–7]. Although some of these hexaferrites have helimagnetic transition temperatures above room temperature, their ME activity is far below the transition temperature (~100 K) because of their low resistivity. By contrast, Z-type $Sr_3Co_2Fe_{24}O_{41}$ has been reported to exhibit high resistivity (~$1.3 \times 10^9$ Ω cm) and a low-field ME effect at room temperature [8]. Electric polarization in Z-type hexaferrites is induced by a magnetic field of several milli-tesla, and the polarization does not vanish during magnetic-field cycling.

Z-type hexaferrites belong to the space group $P6_3/mmc$, and their crystallographic unit cells contain 30 transition metal ions, $Fe^{3+}$ and $Co^{2+}$, with octahedral, tetrahedral, and fivefold coordination [9, 10]. However, the magnetic unit cell in a Z-type hexaferrite can be described by alternate stacks of large and small magnetic moments. The magnetic structure in zero magnetic field consists of a transverse conical structure whose wave vector is expressed as $q = (0, 0, 1)$ in the hexagonal setting [11]. The transverse conical structure can induce electric polarization perpendicular to both the net magnetization and $c$ axis because of the cycloidal component of the magnetic structure. Moreover, the Z-type hexaferrite $Ba_{0.52}Sr_{2.48}Co_2Fe_{24}O_{41}$ exhibits a large ME susceptibility of 3200 ps/m at room temperature, and the magnetization can be controlled by application of an electric field [12]. In ME measurements using single crystals [12], the electric polarization was almost zero in zero magnetic field despite the transverse conical structure formed. This result indicates that the transverse conical domains were randomly distributed in the specimen. However, the magnetic domains in a Z-type hexaferrite have not yet been examined even though several studies on magnetic domains in Y-type and M-type ME hexaferrites have been reported [13–18]. Furthermore, previous magnetization measurements have revealed that Z-type hexaferrites have three magnetic phase transitions and show magnetization along the easy axis ($c$ axis) in a ferrimagnetic phase above the transverse conical phase [8, 12]. These results suggest a high magnetocrystalline anisotropy and the emergence of magnetic bubbles in the ferrimagnetic phase of Z-type $Sr_3Co_2Fe_{24}O_{41}$. Therefore, it is important to determine the magnetic domain structures in both the transverse conical and ferrimagnetic phases of $Sr_3Co_2Fe_{24}O_{41}$.

In this study, we evaluated the magnetic domain structures in magnetoelectric $Sr_3Co_2Fe_{24}O_{41}$ using Lorentz microscopy. We identified the structural characteristics of the magnetic domains in the transverse conical and ferrimagnetic phases and observed the formation of magnetic bubbles in the ferrimagnetic phase. Depending on the orientation of the applied magnetic field with respect to the crystallographic axis, two types of magnetic bubbles were formed by tilting the $c$ axis of the specimen from the magnetic fields. We discuss the evolution of these magnetic domains and the emergence of magnetic bubbles by considering the magnetic anisotropy in the hexaferrite.

## II. EXPERIMENTAL METHOD

Single crystals of $Sr_3Co_2Fe_{24}O_{41}$ (hereafter SCFO) were grown using a flux method ($Fe_2O_3$–$Na_2O$ flux). The crystals had typical dimensions of ~1.3 mm × 1.3 mm × 0.5 mm. The magnetization of the crystals was measured as a function of temperature $T$ and magnetic field $H$ using a vibrating sample magnetometer with an oven option. Thin specimens were prepared for transmission electron microscopy analysis using a focused ion beam or Ar-ion milling after mechanical polishing. High-angle annular dark field scanning transmission electron microscopy (HAADF STEM) and energy-dispersive X-ray spectroscopy (EDS) were performed using a transmission electron microscope equipped with a spherical aberration corrector (JEM-ARM200CF). The STEM–EDS system was equipped with double silicon drift detectors, and the solid angle for the entire collection system was approximately 1.9 sr. The energy of the EDS mappings was set as follows for each element: Fe $K$-edge, 6263–6533 eV; Co $K$-edge, 6789–7069 eV; and Sr $K$-edge, 13968–14358 eV. Lorentz images were obtained using another transmission electron microscope (JEM-2100F) operated at 200 kV, and the objective lens of the microscope was used to apply external magnetic fields perpendicular to the widest face of the thin specimens. The Fresnel method (out-of-focus method) was used to examine the magnetic domain walls [19, 20]. Here we briefly describe the principle of imaging using the Fresnel method (Fig. 1). When the incident electrons pass through magnetic domains, the electrons are deflected by the Lorentz force due to the magnetization, causing the electrons to diverge or converge at the domain walls. If the focus of the imaging lens is overfocused, dark and bright lines appear at the left and right domain wall positions, respectively. The contrast at the domain walls in the overfocused image is reversed if the imaging lens is underfocused. Further, the contrast disappears for an in-focus condition.

## III. RESULTS

### A. Atomic structure

The atomic structure of the single crystal was identified using HAADF STEM and EDS. Figure 2 presents an HAADF image [(a)] and chemical maps [(b)–(e)] at atomic resolution along the *a* direction. Figure 2(f) is a structural model of a Z-type hexaferrite based on Rietveld analysis using neutron diffraction [9, 10]. The model matches well with the STEM and EDS images; therefore, the grown single crystal was confirmed to be a Z-type hexaferrite. Note that although the intensity in HAADF STEM is roughly proportional to the atomic number $Z^2$ [21], some Fe sites are brighter than Sr sites because several Fe atoms are stacked along the *a* axis. The Co atoms showed a clear a preference for the *Me*4(12*k*) site near Sr atoms. Structural analysis of SCFO based on neutron diffraction [9] previously revealed that the Co site occupancies in the transition metal sites *Me*1(2*a*) and *Me*4(12*k*) were higher than those in other transition metal sites. Comparing the EDS maps of Co [Fig. 2(c)] and the structural model [Fig. 2(f)], the *Me*4(12*k*) site is dominantly occupied by Co atoms, which is consistent with the previous structural analysis [9].

### B. Magnetic phase transitions

The magnetization *M* was measured to determine the magnetic phase transitions. Figure 3 shows the magnetization measured at 10 mT as a function of temperature. The temperature dependence of the magnetization shows two clear anomalies at approximately 480 and 680 K. The rise of the magnetization at ~680 K corresponds to a transition from a paramagnetic to a ferrimagnetic phase. The net magnetization was oriented parallel to the *c* axis between 480 and 680 K, whereas below 480 K, the direction of the magnetization was parallel to the *ab* plane.

These results are consistent with those reported in a previous neutron diffraction study [10]. The small jump at ~410 K in the $M$–$T$ curve for $H//c$ indicates a transition from the ferrimagnetic to transverse conical phase [see inset of Fig. 3(a)]. We define the ferrimagnetic phase at 480 K $\leq T \leq$ 680 K as the FM1 phase, the ferrimagnetic phase at 410 K $\leq T \leq$ 480 K as the FM2 phase, and the transverse conical phase below 410 K as the TC phase. Figure 3(b) shows the magnetic-field dependence of the magnetization in the respective phases (at 300 K for TC, 430 K for FM2, and 520 K for FM1). Metamagnetic-like stepwise transitions were only observed in the TC phase when $H$ was applied perpendicular to the $c$ axis, which can be explained in terms of successive changes of the transverse conical axis [22]. The first increase of the magnetization corresponds to the reorientation that causes the helical axis to point in the magnetic-field direction. The second broad increase of the magnetization indicates that the cone angle of the helical structure decreases with increasing magnetic field. Although the magnetization perpendicular to the $c$ axis showed metamagnetism, the magnetization change parallel to the $c$ axis was less pronounced because the $ab$ plane is the easy plane below 480 K. At 430 K (FM2), the saturation magnetic field for $H//c$ was smaller than that for $H \perp c$, indicating that the easy axis of magnetization was perpendicular to the $c$ axis. Conversely, the magnetic easy axis became parallel to the $c$ axis in the FM1 phase at 520 K, which is consistent with the temperature dependence of the magnetization.

**C. Examination of magnetic domain structures**

Next, we investigated the effect of these magnetic phase transitions on the magnetic domain structures using the Fresnel method. Figure 4 shows the observed magnetic domain structures projected in the directions perpendicular [Figs. 4(a)–(c)] and parallel [Figs. 4(d)–(e)] to the $c$ axis in the respective phases. To examine the formation process of the magnetic domains, the observation was performed during cooling and heating for the data presented in Figs. 4(a)–(c) and 4(d)–(e), respectively, because no magnetic domain was observed in Figs. 4(c) and 4(d). This process eliminated the possibility that the observed magnetic domains were formed during the thinning stage of the specimen preparation. Figures 4(a) and 4(d) present magnetic domain images projected in directions perpendicular and parallel to the $c$ axis, respectively, in the TC phase. Basing on these results, one can conclude that the TC phase contains numerous fine stripe-shaped domains with typical domain widths of ~20 nm and that the domain boundary is perpendicular to the $c$ axis. Because the stripe-shaped domain size is small and the spatial resolution is not high in the Fresnel image because of the defocus, the domain walls (the bright and dark contrast) appear to be combined. The transverse conical domains in Fig. 4(a) indicate that the conical axes in the domains change their directions at a nanometer scale because the bright and dark contrast depend on the direction of the macroscopic magnetization in each domain. In the FM2 phase [Figs. 4(b) and (e)], 180° magnetic domains with their boundaries perpendicular to the $c$ axis were formed, and the magnetization direction was perpendicular to the $c$ axis. This direction is consistent with the easy axis revealed by the magnetization measurements displayed in Fig. 3. The 180° magnetic domain width was ~400 nm. The left and right magnetized domains were almost equally populated, and these domains were separated by magnetic domain walls showing bright or dark contrast. Note that the type of domain wall (Bloch or Néel wall) cannot be determined from the image because the contrast at the domain walls was caused by the magnetization in the 180° domains, and no difference is observed between Bloch and Néel walls [23]. During the phase transition from the FM2 to TC phase, some domain structures remained unchanged; however, the fine striped domains disappeared in the FM2 phase (as observed by comparing the areas in Figs. 4(a) and 4(b) marked by white and black lines). The contrast of the stripe-shaped domains was caused by the convergence or divergence of the electrons deflected by the magnetizations between the domains. Thus, we consider the directions of magnetization in the stripe-shaped domains to be opposite those in the FM2 phase, as illustrated in Figs. 4(g) and (h). In the FM1 phase, further drastic change in the magnetic domain structure occurred. Unlike in the TC and FM2 phases, no magnetic contrast was observed in the image projected in the direction perpendicular to the $c$ axis [Fig. 4(c)], whereas pairs of bright and dark contrast were formed in the FM1 phase [Fig. 4(f)]. This result indicates that up- and down-magnetized domains ($M//c$) were formed and that Bloch walls were located between the domains in the FM1 phase, as illustrated in Fig. 4(i). A Bloch wall can be visualized as a pair of bright and dark contrast when the magnetization in the domains is parallel to the incident beam (the out-of-plane magnetization), and the Bloch wall has an in-plane component [24]. The magnetization was parallel to the $c$ axis, which agrees with the macroscopic magnetization measurements [Fig. 3(a)]. The wavy patterns in Fig. 4(f) demonstrate that the magnetic anisotropy is equivalent within the $ab$ plane.

**D. Observation of magnetic bubbles**

The magnetic-field dependence of the magnetization [Fig. 3(b)] showed a strong uniaxial magnetic behavior in the FM1 phase. Recent Lorentz microscopy experiments demonstrated that magnetic bubbles are formed in magnetic materials with a high magnetocrystalline anisotropy ($K_u$) [25–27]. Therefore, magnetic bubbles would be expected to form in the FM1 phase upon applying a magnetic field to a thin specimen along the $c$ axis with the widest face perpendicular to the $c$ axis. The results of such an experiment are displayed in Fig. 5. Note that the magnetic-field direction was downward with respect to the magnetization in these experiments. When a magnetic field of 95 mT was applied at 520 K along the $c$ axis, the down-magnetized domains expanded and the up-magnetized domains shrank. As a result, circularly rotating magnetic domain walls were created, as observed in Fig. 5(a). The magnetic bubbles were all type I, in which the magnetic domain walls were magnetized continuously clockwise or counterclockwise, as shown in Fig. 5(d). The magnetic domain structure had the same topological number [$N = \iint \frac{1}{4\pi} \boldsymbol{n} \cdot \left(\frac{\partial \boldsymbol{n}}{\partial x} \times \frac{\partial \boldsymbol{n}}{\partial y}\right) dS = 1$] as that in magnetic skyrmions, where $\boldsymbol{n} = \boldsymbol{M}/|\boldsymbol{M}|$ is the unit of the magnetization vector [14]. The diameter of the magnetic bubbles was approximately 300 nm, which is a typical size for magnetic bubbles [28, 29].

We also investigated the relation between the easy axis and magnetic-field direction. Figures 5(b) and 5(c) show the magnetic-field dependence of the magnetic domains when the $c$ axis of the specimen was tilted by 5° from the magnetic-field direction. The magnetic domains were pinched off from the edges to create type-II magnetic bubbles with increasing magnetic fields. The type-II magnetic bubble consisted of two parallel domain walls with a pair of Bloch lines, as illustrated in Fig. 5(e). Unlike the type-I structure, the type-II structure did not have a topological number ($N = 0$). The type-II magnetic bubble had a higher domain wall energy than the type-I magnetic bubble because the Bloch line contained in-plane magnetic components. The energy loss was compensated by the Zeeman energy caused by the in-plane external magnetic field, and thus, the type-II magnetic bubbles were stabilized in this configuration. The change in the type of magnetic bubbles achieved by tilting a specimen was also recently reported in a uniaxial magnet, $BaFe_{12-x-0.05}Sc_xMg_{0.05}O_{19}$ [13].

## IV. DISCUSSION

### A. Transverse conical domain structure

The fine stripe-shaped domain structures observed at 300 K (TC) developed when the specimen underwent a phase transition from the FM2 phase with the 180° domains. Therefore, we consider these fine domains to be associated with transverse conical domains. Soda and coworkers reported that the propagation vector of the transverse conical structure is $q = (0, 0, 1)$ based on their neutron diffraction study [11]. Thus, the helical period is ~5.2 nm (the lattice constant of the $c$ axis). The fine irregular domains of ~20 nm [Fig. 4(a)] correspond to approximately four periods of the unit cell. We believe that this short coherence length of the transverse conical domains originates from the microscopic mechanism that stabilizes the helical structure because the observed fine domains differ from helical domains induced by the DM interaction. In chiral magnets, because of the DM interaction, helical domains are long wavelength and continuous throughout a specimen even when defects or edges are present [30, 31]. This finding is observed because the DM interaction depends on the crystal structure, which determines the helical structure. Unlike in chiral magnets, in Z-type hexaferrite, the transverse conical structure is caused by the superexchange interaction in an Fe–O–Fe bond [4, 11]. The Fe atoms are located across the boundary between the blocks that form the large and small magnetic moments, and Ba and Sr ions are located near the boundary. Therefore, the interaction between the blocks along the $c$ axis is considered small compared with the ferromagnetic exchange interaction within the $ab$ plane. Consequently, the transverse conical domains have a short period along the $c$ axis, whereas the interaction perpendicular to the $c$ axis is long, producing the irregular stripy domains observed in Fig. 4(a).

We also compare the helical domains in SCFO with those in MnP because MnP has similar characteristics to SCFO. MnP is ferromagnetic between 47 and 291 K and exhibits helimagnetism below 47 K. The helimagnetism is thought to be realized through several superexchange interactions, and the helical period is $9a \sim 5.3$ nm ($a$ is the lattice constant) [32], although the DM interaction has been reported to modify the spin configuration [33]. The ferromagnetic domains are 180° domains separated by the Bloch walls in the ferromagnetic phase [34]. The chiral domains (left- and right-handed helical domains) of MnP have been previously examined using polarized neutron diffraction topography by Patterson and coworkers [35]. The chiral domain patterns in the helical phase had irregular striped shapes, and the domains were elongated along the easy axis. They reported that the first-order ferro-helimagnetic transition of the stripe-type domains started in the Bloch walls. The nucleation in the helical phase starts in a Bloch wall, and its growth occurs mainly along the easy axis to minimize its magnetostatic energy. Although the type and size of helical domains differ, we speculate that the same discussion can be applied to SCFO. That is, the helical domains grow along the $ab$ plane (the easy plane), and thus, the striped domains are formed to reduce the magnetostatic energy, as observed in Fig. 4(a). We note that the helical domains were nanosized in this study because the domain size tends to decrease in a thin foil compared with that in the bulk to minimize the magnetostatic energy (the magnetic flux density generated from the specimen to vacuum). Furthermore, a recent resonant X-ray diffraction study of a Y-type hexaferrite suggested that stripe-type domains with flat domain boundaries extend perpendicular to the $c$ axis [17]. The origin of this behavior is considered to be the first-order ferro(fan)-helimagnetic transition starting in Bloch walls, as explained in the domains of MnP. Considering the similarity of the crystal and magnetic structures of Y-type and Z-type hexaferrites, the same phenomena could be possible in the transition from the ferrimagnetic to helical phases in SCFO.

**B. Formation of magnetic bubbles**

In the FM1 phase, we observed magnetic bubbles whose type (type I or II) depended on the magnetic-field direction. To form magnetic bubbles, the anisotropy field $H_k$ (= $2K_u/M_s$) must be larger than the demagnetizing field $4\pi M_s$, where $K_u$ is the uniaxial anisotropy coefficient and $M_s$ is the saturation magnetization [25, 26, 29]. The anisotropy field $H_k$ is defined as the critical value at which the magnetizations parallel and perpendicular to the $c$-axis are the same, as indicated by the arrowheads in Fig. 3(b). If this requirement is not satisfied, the magnetization points parallel to the surface of a thin-film sample because of the magnetic shape anisotropy; the magnetocrystalline anisotropy does not determine the magnetization direction. Figure 3(b) shows that the anisotropy field $H_k$ was 0.47 T (4700 G) and that the saturation magnetization $M_s$ was 13 $\mu_B$/f.u. (179 G) at 520 K. Thus, the requirement for the formation of magnetic bubbles $Q = H_k/4\pi M_s = K_u/2\pi M_s^2 \sim 2.0 > 1$ was fulfilled in the FM1 phase. Here, the ratio of the anisotropy and demagnetizing fields $Q$ is called the quality factor. The uniaxial anisotropy coefficient $K_u$ (=$H_kM_s/2$) obtained from $H_k$ and $M_s$ was $1.4 \times 10^4$ J/m$^3$, which is relatively smaller than that of other magnetic oxides exhibiting magnetic bubbles [28]. However, the magnetization $M_s$ was also smaller because of the ferrimagnetism in SCFO, resulting in satisfaction of the requirement. Note that magnetic bubbles were not produced in the FM2 phase. In the FM2 phase, $H_k$ was also large, as indicated by the black arrowheads in Fig. 3(b). However, the magnetization can rotate within the *ab* plane in the FM2 phase because the easy plane is the *ab* plane. Therefore, the magnetic domains were 180° domains in the FM2 phase because of the magnetic shape anisotropy, as observed in Fig. 4(b). Thus, the striped magnetic domains observed in the FM1 phase were not formed in the FM2 phase.

In addition, the magnetic bubbles observed in this paper are the typical structure whose magnetizations gradually point in the out-of-plane direction from the wall position. However, it has been reported that another magnetoelectric hexaferrite (Sc-substituted M-type hexaferrite) forms magnetic bubbles with helicity reversals [13]. The bubble structures were observed in M-type hexaferrite with small uniaxial anisotropy because of the Sc substitution ($Q \sim 1$) and thin-film thickness ($h \sim 30$ nm). In the study, the helicity reversals were reproduced by numerical simulation based on the Garel–Doniach model [36], which describes ferromagnetic states (stripes and bubbles) in thin films using two parameters (the uniaxial anisotropy and film thickness). In our study, the quality factor ($Q = 2.0$) was higher than that of M-type hexaferrite and the thickness was likely thicker than that of the previous study, which explains the differences between the bubble structures.

## V. SUMMARY

In summary, we studied the evolution of magnetic domain structures in a ME hexaferrite, $Sr_3Co_2Fe_{24}O_{41}$, forming various magnetically ordered phases by changing the temperature and magnetic fields. Using Lorentz microscopy, we observed that this hexaferrite contained characteristic domain structures such as fine striped magnetic domains in a ME transverse conical phase as well as magnetic bubbles upon application of magnetic fields in a high-temperature ferrimagnetic phase. In addition, we revealed the uniaxial behavior and change of the easy axis in the ferrimagnetic phases based on magnetization measurements. Z-type hexaferrites exhibit high resistivity, and thus, the electric-field-driven motion of magnetic bubbles, which depends on the topological number ($N = 0$ or 1), is expected [37, 38]. Our findings demonstrate the versatility of a room-temperature ME Z-type hexaferrite for spintronic memory devices that use the magnetic bubble properties and magnetoelectric effect.

## ACKNOWLEDGEMENTS

We thank K. Shimada and K. Harada (*RIKEN* CEMS) for the sample preparations using the focused ion beam. This work was partially supported by Grant-in-Aid (Nos. 17H01143, 16H03833 and 15K13306) from the Ministry of Education, Culture, Sports, Science and Technology (MEXT), Japan.

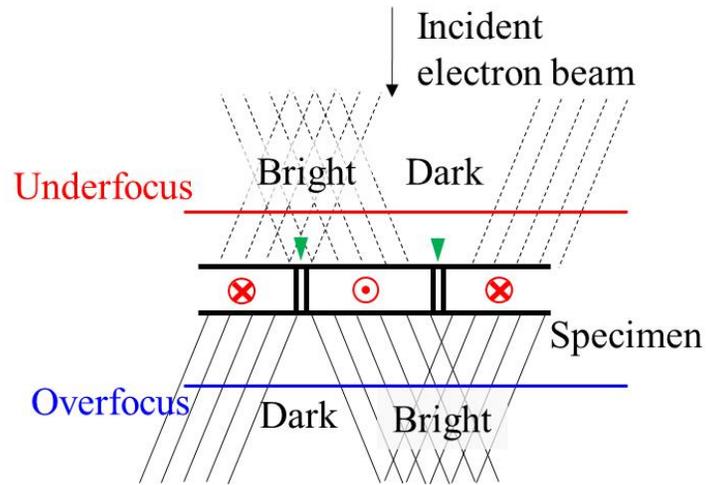

FIG. 1. (a) Schematic of imaging in the Fresnel method. The red marks represent the magnetization directions in the respective domains. The green arrowheads mark the positions of the domain walls.

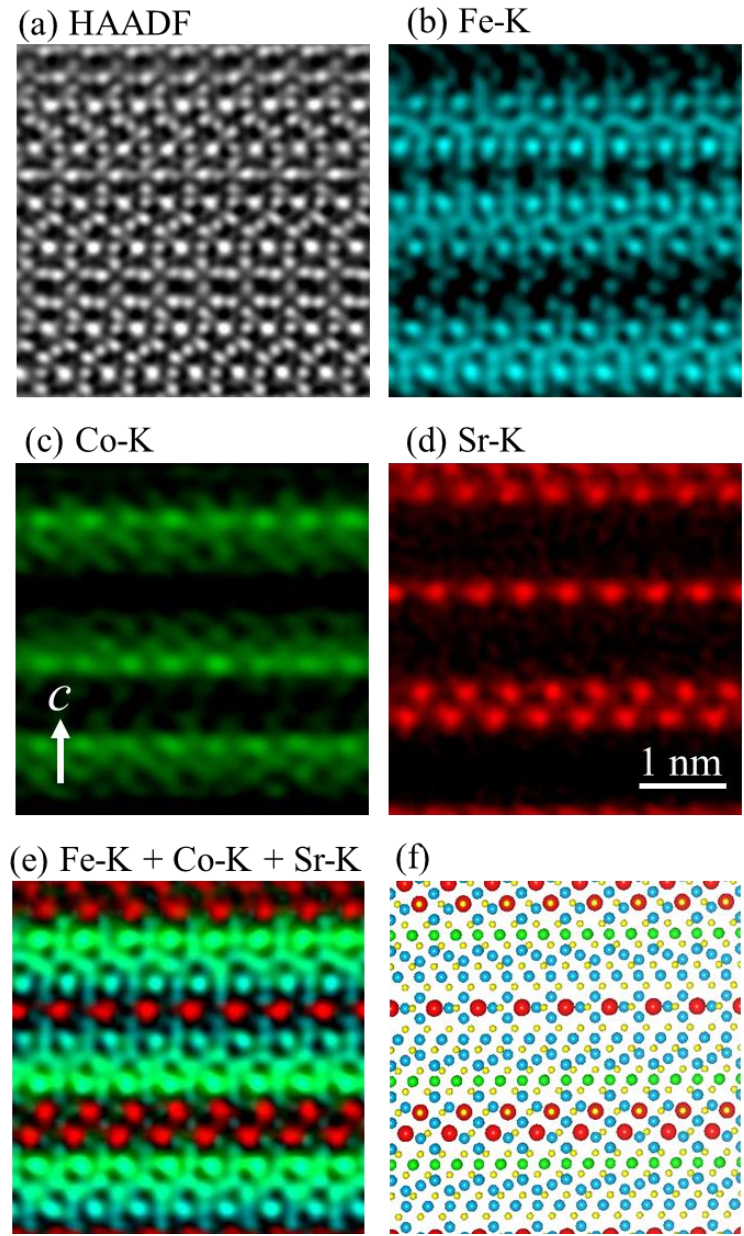

FIG. 2. (a) HAADF STEM image of SCFO. Atomic-resolution EDS chemical maps showing the (b) Fe-$K$, (c) Co-$K$, and (d) Sr-$K$ edges. (e) Merged image of (b)–(d). The incident electron beam is along the $a$ axis, and the white arrow in panel (c) shows the direction of the $c$ axis. (f) Schematic of the crystal structure of SCFO along the $a$ axis. The blue, red, and yellow balls represent Fe, Sr, and O atoms, respectively. The metal sites, $Me4(12k)$, are marked in green.

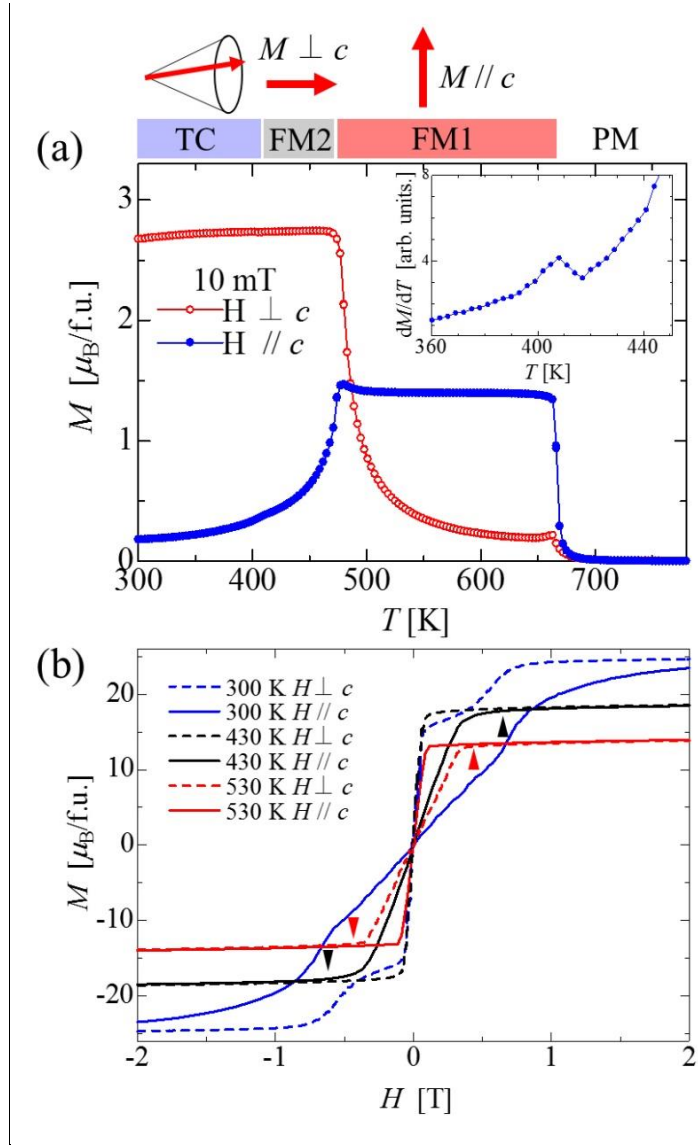

FIG. 3. (a) Temperature dependence of the magnetization parallel and perpendicular to the $c$ axis of SCFO. The inset shows the derivative of the magnetization (d$M$/d$T$) at approximately 400 K. Schematics of the magnetization directions are indicated by red arrows. PM, FM, and TC represent the paramagnetic, ferrimagnetic, and transverse conical phases, respectively. (b) Magnetic-field dependence of the magnetization parallel and perpendicular to the $c$ axis at various temperatures. The arrowheads mark the anisotropy field $H_k$.

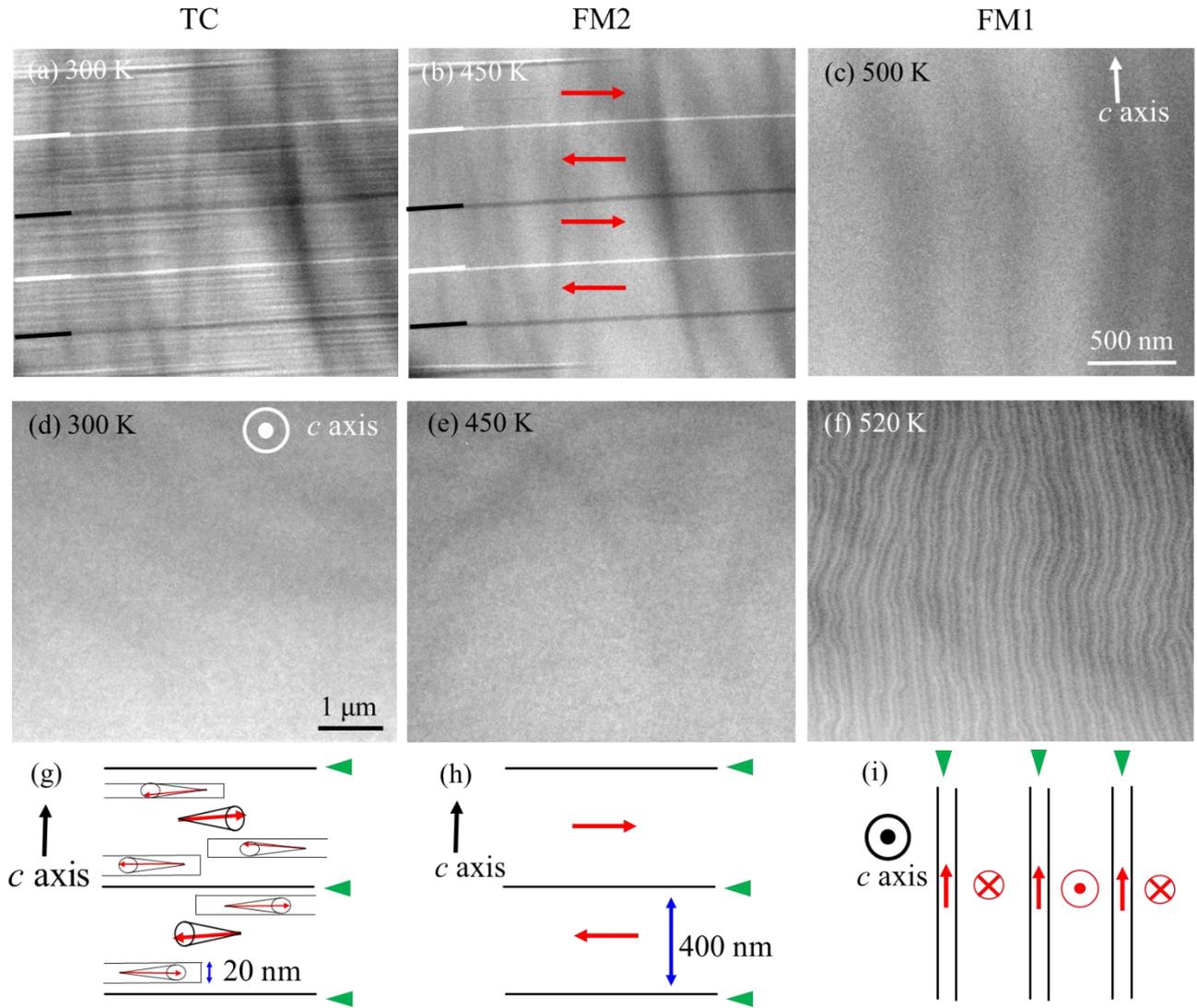

FIG. 4. Magnetic domain structures in the TC [(a) and (d)], FM2 [(b) and (e)], and FM1 [(c) and (f)] phases observed using the Fresnel method. The widest faces of the thin samples used for these experiments were parallel [(a)–(c)] and perpendicular [(d)–(f)] to the $c$ axis. The white and black lines in (a) and (b) represent bright and dark contrast, respectively. The defocus value was −5 μm. (g)–(i) Schematics of the magnetic domains in panels (a), (b), and (f), respectively. The arrowheads mark the domain walls in the FM1 and FM2 phases. The red arrows represent the magnetization directions in the respective domains and domain walls.

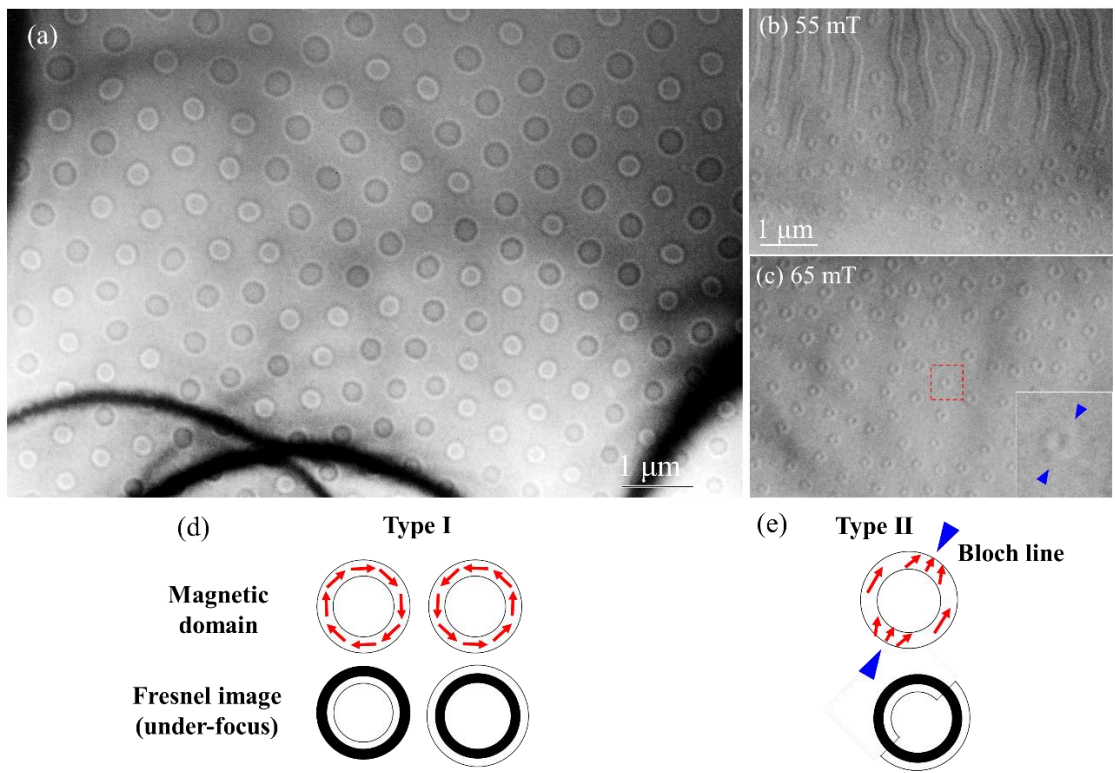

FIG. 5. (a) Fresnel image of clockwise and counterclockwise type-I magnetic bubbles at 95 mT and 520 K (FM1 phase). The plane is perpendicular to the $c$ axis, and the magnetic field was applied parallel to the $c$ axis. The defocus value was −2 μm. (b), (c) Formation process of type-II magnetic bubbles at 520 K (FM1 phase). The specimen (the $c$ axis) was tilted by 5° from the magnetic-field direction. The defocus value was −5 μm. The inset of (c) presents a magnified image of the magnetic bubble marked by a red dashed rectangle. Schematics of magnetic bubbles of (d) type I and (e) type II and their corresponding Fresnel images in an underfocused condition. The blue arrowheads mark the positions of Bloch lines, where the chirality of a domain wall is reversed.